\documentclass[twocolumn,showpacs,preprintnumbers,prb,amsmath,amssymb]{revtex4}
\usepackage{graphicx}
\usepackage{dcolumn}
\usepackage{bm}
\usepackage{longtable}
\usepackage{array}
\usepackage{color}

\begin{document}

\title{Coulomb interaction signatures in self-assembled lateral quantum dot molecules}
\author{Xinran. R. Zhou$^{1}$}
\author{Jihoon. H. Lee$^{2,3}$}
\author{Gregory. J. Salamo$^{3}$}
\author{Miquel. Royo$^{4}$}
\author{Juan. I. Climente$^{4}$}
\author{Matthew. F. Doty$^{1}$}
\email{doty@udel.edu}

\affiliation{$^{1}$Dept. of Materials Science and Engineering, University of Delaware, Newark, DE 19716, USA}
\affiliation{$^{2}$School of Electronics and Information, Kwangwoon University, Nowon-gu, Seoul 139-701, South Korea}
\affiliation{$^{3}$Institute of Nanoscale Science and Engineering, University of Arkansas, Fayetteville, Arkansas 72701, USA}
\affiliation{$^{4}$Departament de Quimica Fisica i Analitica, Universitat Jaume I, E-12080, Castello, Spain}

\date{\today}

\begin{abstract}
We use photoluminescence spectroscopy to investigate the ground state of single self-assembled InGaAs lateral quantum dot molecules. We apply a voltage along the growth direction that allows us to control the total charge occupancy of the quantum dot molecule. Using a combination of computational modeling and experimental analysis, we assign the observed discrete spectral lines to specific charge distributions. We explain the dynamic processes that lead to these charge configurations through electrical injection and optical generation. Our systemic analysis provides evidence of inter-dot tunneling of electrons as predicted in previous theoretical work.
\end{abstract}

\maketitle
\section{Introduction}\label{intro}
Self-assembled semiconductor quantum dot molecules (QDMs), a system composed of at least two closely-spaced quantum dots (QDs), have attracted broad attention for both studies of fundamental physics and development of next-generation optoelectronic devices.\cite{Doty2010, Liu2010, Vamivakas2010}. QDs are often called "artificial atoms" because of their discrete energy states, which can be controlled by QD geometry during the growth process\cite{Wang2009}. QDMs are interesting for fundamental studies of the interaction between confined charges and spins and evolution of molecular states in solid state system. QDMs are also of interest as a component of optoelectronic devices because the molecular coupling can be controlled by electric and magnetic fields. Consequently, QDMs are a promising material for single spin/charge optoelectronic devices \cite{Kim2010b} including quantum computing devices \cite{Economou2012}.

In the past decade, most research related to QDMs has focused on vertically-stacked QDMs (VQDMs)\cite{Krenner2005, Stinaff2006, Doty2006, Doty2009, Ortner2005}. In a VQDM, two or more adjacent dots separated by thin barrier(s) are stacked one by one along the growth axis\cite{Ledentsov1996}. Using well-established epitaxial growth protocols, geometrical properties such as height, barrier thickness and relative position of the QDs can be all precisely controlled. By applying an electric field along the growth axis, the energy levels of the QDs can be simultaneously tuned relative to one another and to a doped substrate. In order to independently control the charge and tunneling, QDMs comprised of pairs of laterally separated QDs arranged along axes perpendicular to the growth direction (LQDMs) are of interested\cite{Lee2006, Liang2006, Liang2008}. Parallel and independent control of coupling and charge manipulation in multiple QDMs is a prerequisite for scaling up and building optoelectronic devices that use the QDs as bit registers.

Coherent interactions between neighboring QDs lead to the formation of molecular-like delocalized states in LQDMs. Peng and Bester have used  atomistic empirical pseudopotential calculations to calculate the energies of excitons with different charge configurations in InGaAs LQDMs under lateral electric fields\cite{Peng2010}. In the recent publication of Royo et. al, the optical resonance of neutral and charge excitons in LQDMs was simulated as a function of inter-dot distance\cite{Royo2011}. Both of these papers predicted sensible signatures of tunnel coupling in charged states. However, the experimental evidence of controllable inter-dot tunnel coupling in LQDMs\cite{Zhou2011, Munoz2011, Heldmaier2012, Beirne2006} remains indirect. The center-to-center distance between a QD pair in a LQDM is about ten times larger than the separation in a VQDM, the tunneling strength is expected to be significantly weaker. Furthermore, tuning the tunnel coupling while deterministically controlling the total electric charge individual QDs in LQDMs is still a challenge.

We present a systematic analysis of the photoluminescence (PL) emission of self-assembled InAs LQDMs under a voltage applied along the growth direction. We observe a series of discrete PL lines with distinct energy shifts with the increasing electric field. We assign these discrete lines to specific charge configurations using a combination of theoretical modeling and analysis of the formation dynamics. We compare the spectral signatures of LQDMs in which the two QDs have similar and different confined energy states. The measured spectral shifts support the conclusion that inter-dot electron tunneling is present in trion states.

\section{Experimental}\label{Expt}
\begin{figure}[htb]
\begin{center}
\includegraphics[width=8.0cm]{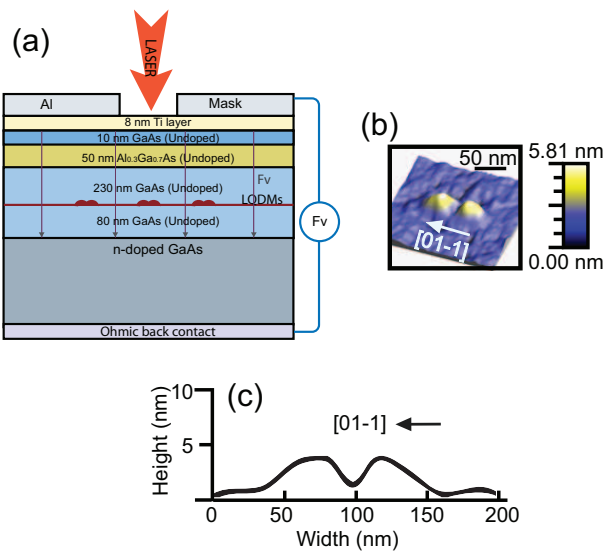}
\caption{(Color Online) a) The layout of the self-assembled LQDMs in a Schottky diode. This structure applies a electric field strictly along the growth direction. b) AFM image and c) cross-sectional profile of single LQDM grown in the same condition as the sample we studied. \label{exp}}\end{center}
\end{figure}

The self-assembled InAs/GaAs LQDMs sample we studied was grown by Salamo's group of the University of Arkansas by solid-source molecular beam epitaxy (MBE)\cite{Lee2010}. The QDs were grown on an n-doped GaAs [1 0 0] substrate. The first step is growing single InAs QDs by Stranski-Krastanov mode on the undoped GaAs surface. Then, the dome-shaped QDs are partially capped by a 10-ML thick GaAs layer. During an in-situ annealing at 480 C, the InAs diffuses anisotropically along the [0 1 -1] direction of GaAs surface and evolves into the InAs QD pairs. An AFM image of uncapped LQDMs are shown in Fig.~\ref{exp} (a). A cross-sectional profile of one single LQDM is shown in Fig.~\ref{exp}(b). For optical characterization and devices, GaAs and AlGaAs are deposited to cap the QDMs underneath. (See Fig.~\ref{exp}(b)). Devices are processed with Ohmic back contact and Ti top contact to create a Schottky diode structure with a 19.2 kV/cm built-in voltage in the LQDMs sample. With an increase of the applied voltage, the confined QD energies will drop toward the Fermi level set by the n-type doping.

The areal density of the LQDMs is about 30 LQDMs per $\mu$m$^2$. In order to acquire the spectra of single LQDMs and control the bias along the growth direction of LQDMs, an Al Schottky contact with 1 $\mu$m gap is applied on the top surface of the sample by electron-beam deposition. A semi-transparent Ti thin layer is also applied in between of the sample and Al layer to force the electric field to be strictly along the growth direction. We refer to this as a vertical electric field.

The sample was mounted in a close-cycle cryostat cooled to 12 K and excited by a linearly polarized 870 nm laser for PL spectroscopy. The emitted photoluminescence was collected by a high numerical aperture objective and analyzed with a liquid nitrogen cooled charge-coupled device camera with 70 ueV spectral resolution.

\section{Photoluminescence of neutral excitons and positive trions in LQDMs}\label{positive}
In our past work\cite{Zhou2011}, we observed the coulomb shift of ensemble and single LQDMs PL in both ground and first excited shells under the electric field. The results validate that ground electron and hole states of a LQDM are localized to individual QD while the first and higher excited electron states are delocalized over the entire LQDM. In this paper, we focus on the coulomb interactions in the localized ground states of a single LQDM in order to understand the effect of charge occupancy.

\begin{figure}[htb]
\begin{center}
\includegraphics[width=8.0cm]{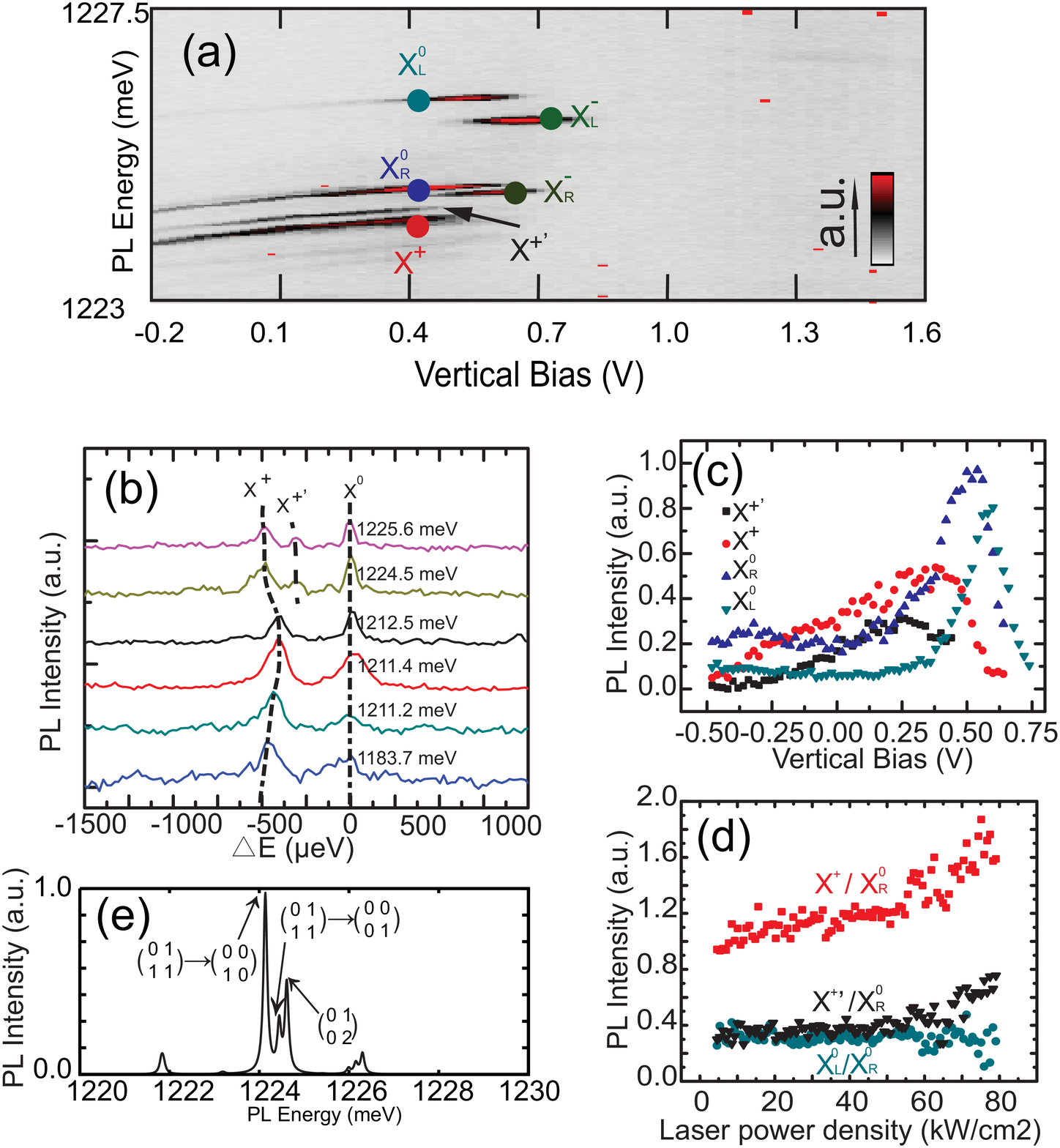}
\caption{(Color Online) (a)PL spectrum of the ground states transitions in LQDM 1 as a function of vertical voltage. (b) PL spectra of $X^{0}$ and $X^{+}$/$X^{+}$' in 6 different LQDMs. Lines are guides to the eye. Energies are plotted relative to the emission of the $X^{0}$ state, with the absolute energy of the $X^{0}$ state indicated by the inset text. (c) Intensities of PL lines $X^{0}_{R}$, $X^{+}$/$X^{+}$' and $X^{0}_{L}$ as a function of vertical voltage. (d) Intensity ratios of PL lines as a function of laser power densities under 0.41 V vertical voltage. (e) Theoretical modeling results of the energies and intensities of positive trion emissions in LQDMs based on the energies of neutral exciton emissions. \label{positiveF}} \end{center}
\end{figure}

The color map in Fig.~\ref{positiveF}(a) presents the discrete PL signature of a single LQDM (LQDM 1) structure, as a function of applied vertical electric field. The x axis is the electric bias applied along the growth direction of the LQDMs while the y axis indicates the PL energy of the discrete PL lines. The color shows the intensities of PL emission, as represented in the inset of Fig.~\ref{positiveF}(a). The energies of this set of PL lines indicate that the PL emission is from the ground states of the LQDM\cite{Zhou2011}. Typically, the two QDs, which comprise the LQDM will be slightly different in energy levels. Although we cannot assign PL emission to the right or left QD, we simplify the discussion by always assigning the low-energy PL emission to the right QD and the high-energy PL to the left QD.

At negative bias, three PL lines ($X^{0}_{R}$, $X^{+}$ and $X^{+}$') are observed (at 1223.7, 1223.9 and 1224.2 meV, respectively). A fourth PL line $X^{0}_{L}$ is also observed in 1225.7 meV with relatively weak intensity. With increasing applied voltage, these four lines show identical stark energy shifts and distinct changes in PL intensities (See Fig.~\ref{positiveF}(c)). As the applied voltage reaches 0.58 V (0.66 V), line $X^{0}_{R}$ (line  $X^{+}$/$X^{+}$') is turned off. Parallel PL lines with approximately 0.5 meV separation, such as line $X^{+}$/$X^{+}$' and $X^{0}_{R}$, are a characteristic signature in the PL of single LQDMs. Figure. ~\ref{positiveF}(b) presents a survey of the parallel PL lines in six LQDMs in our sample. In all of these examples, the energy separation between the two dominant PL lines are between 0.4 and 0.5 meV. This pair of PL lines with small and constant energy separation could originate from either 1) recombination involving the energy levels of two different QDs or 2) recombination of one single QD with the presence of different number of spectator charge in the LQDM. As indicated by the state labels, we assign the pair of lines to the $X^{0}$ and $X^{+}$ charge configurations of a single QD within the LQDM. We now justify this assignment.

In the n-type Schottky diode, the electron energy states of QDs are higher than the doped Fermi level under negative bias. Consequently, electrons participating in the PL emission must be optically generated. Under moderate excitation laser intensity, the number of electrons occupied a single LQDM during the PL emission should be no more than one. Therefore, PL lines that show up at negative bias are assigned to either neutral exciton ($X^{0}$) or positive trion ($X^{+}$) states. Due to the anisotropic self-assembly growth mechanism of the LQDMs sample, it is unlikely that the energy difference between two neighboring QDs would consistently be 0.4 to 0.5 meV. This suggests that the two consistently observed pair of PL lines separated by 0.5 meV should not be assigned to two separate QDs.

To support the assignment that the high energy PL is from neutral exciton emission and the low energy PL line is from positive trion emission, we look at the electric field and laser power dependence of the intensities of these lines. The peak intensities from PL emission of line $X^{+}$, $X^{+}$', $X^{0}_{R}$ and $X^{0}_{L}$ are plotted as a function of vertical voltage in Fig.~\ref{positiveF}(c). The intensities of line $X^{+}$ and $X^{+}$' show the same nearly-linear dependence on the voltage and reach their maximum at 0.25V before dropping to zero. Line $X^{0}_{R}$ and $X^{0}_{L}$ both have low intensity until a certain voltage and then gain intensity abruptly. The intensity of PL emission from $X^{0}$ and $X^{+}$ depends on the efficiency of two processes. First, the $X^{+}$ configuration is more likely to form at low bias, when it is relatively easy for an optically generated electron to tunnel out of the LQDM, leaving behind an excess optically-generated hole, thereby increasing the probability that the $X^{+}$ will form. Second, larger electric fields drive e-h pairs to separate and therefore weaken the PL emission of both X0 and X+. As shown in Fig.~\ref{positiveF}(c), line $X^{0}_{R}$ and $X^{0}_{L}$ sharply increase in intensity as line $X^{+}$ and $X^{+}$' begin to get weaker, near 0.25 V. This is the turning point at which it is no longer favorable for the electron to tunnel out of the QD and the emission of $X^{0}$ is therefore enhanced. This analysis supports the assignments of the PL lines in Fig.~\ref{positiveF}(a).

Integrated PL intensities of the lines in Fig.~\ref{positiveF}(a) as a function of laser power density further support this assignment, as shown in Fig.~\ref{positiveF}(d). The ratio of PL intensities of line $X^{0}_{R}$ and $X^{0}_{L}$ remain constant with increasing laser power. In contrast, line $X^{+}$ shows increasing intensity relative to line $X^{0}_{R}$ with increasing laser power. Compared with holes, electrons have a smaller effective mass and are able to tunnel out of QDs more rapidly. The formation of the $X^{+}$ necessarily requires two photons. One photon generates an electron hole pair from which the electron tunnels out, leaving a hole behind. The second photon generates an additional electron hole pair, allowing recombination of an electron hole pair in the presence of the additional hole. Consequently, formation of the $X^{+}$ state requires more photons than formation of $X^{0}$ state. The superliner laser power dependence of intensities of lines $X^{+}$/$X^{+}$', relative to $X^{0}$, indicates that these PL line should be assigned to the positive trion states. This laser power dependence allows us to assign line $X^{0}_{R}$ to emission involving neutral excitons. The end result of the analysis based on the data shown in Fig.~\ref{positiveF}(c) and (d) allows us to conclude that line $X^{0}_{R}$ and line $X^{0}_{L}$ are both from the recombination of neutral excitons (X0) while line $X^{+}$ and $X^{+}$' are the PL recombination with a spectator hole ($X^{+}$). We believe that $X^{0}_{L}$ is weaker than $X^{0}_{R}$  because it is energetically favorable for the electron to relax to the lower energy (right) QD.

The red shift of $X^{+}$ relative to $X^{0}$ differs from the case of single QDs\cite{Finley2004,Dalgarno2008} and VQDMs\cite{Doty2008}, where blue shifts of the $X^{+}$ state are typically observed. This red shift is one of the distinct PL properties of LQDMs that has been predicted by both pseudopotential\cite{Peng2010} and effective mass\cite{Royo2011} modeling.
In the $X^+$ initial state, one electron-hole pair sits in the right QD and the remaining hole in the left QD. If the QDs have very different energies, tunnel coupling is negligible and the electron-hole pair only feels the nearby hole as a static electric field. This reduces the exciton binding energy leading to a Coulomb-induced red shift. If the QD energies are similar enough, the electron may tunnel between the two QDs. This leads to an additional tunnelling-induced red shift of the $X^+$ emission.\cite{Royo2011} In order to determine whether the red shift of $X^+$ in Fig.~\ref{positiveF}(a) is due to Coulomb coupling or to tunnel coupling, further information is needed. The presence of a splitting between $X^+$ and $X^{+'}$ lines gives us a hint that can be interpreted from comparison with theoretical calculations in order to definitely assign these states to specific charge configurations.


The LQDM PL spectra are computed as a function of charge configurations using the model and material parameters described
in Ref.~\onlinecite{Royo2011}. We take a typical distance between QD centers of $d=35$ nm, and parabolic confinement frequencies
of $\hbar \omega_L=25$ meV and $\hbar \omega_R=23.5$ meV, consistent with the experimental QD sizes and the energy splitting
between $X^0_L$ and $X^0_R$ lines. Since the hole is more confined than the electron, we assume the characteristic lengths to
be related by $l_h=0.6\,l_e$.  The energy gap is taken so as to fit the energy of the $X^{0}_{R}$ line. The resulting PL spectrum
for $X^+$ is plotted in Figure~\ref{positiveF}(e).

To clearly indicate the charge configurations, we use the notation $\begin{pmatrix}e_{L}&e_{R}\\h_{L}&h_{R}\\\end{pmatrix}$, where $e_{L}$ ( $e_{R}$)) indicates the number of electrons in the left (right) QD. Similarly for holes. For example $\begin{pmatrix}0&1\\1&1\\\end{pmatrix}$ describes a LQDM with one electron in the right QD and one hole in each QD.

As shown in Figure~\ref{positiveF}(e), there are two peaks that originate in PL emission from the initial state $\begin{pmatrix}0&1\\1&1\\\end{pmatrix}$. The presence of one hole in each QD breaks the Coulomb attraction that binds the electron to a single QD and electron tunneling leads to the formation of a delocalized electron state. The delocalized electron can recombine with both holes, in either the left or right QD, and emit PL at slightly different energies. Compared with the direct recombination involving only electron and hole in the right QD, the indirect $X^{+}$ recombination exhibit significant lower intensity because relatively small probability of tunneling of electron. This allows us to assign X+ to the direct recombination of (01,11) and X+' to the indirect recombination. The modeling result is also in good quantitative agreement with the data in Fig.~\ref{positiveF}(a).

\begin{figure*}[htb]
\begin{center}
\includegraphics[width=16.0cm]{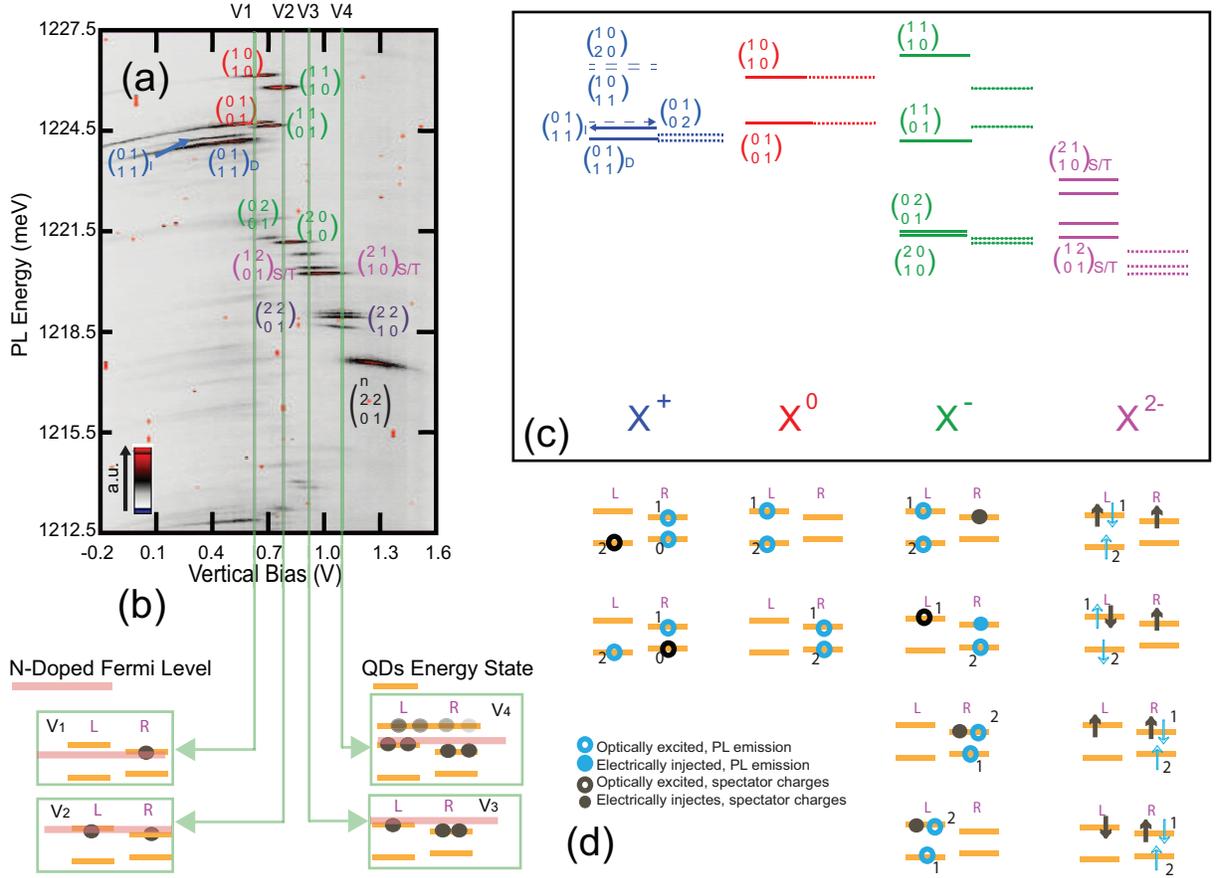}
\caption{(Color Online) (a)Full PL spectra of the ground states transition mapping as a function of vertical voltage measured for LQDM 1. (b) Electrical charging sequence as the applied voltage increases. (c) Comparison between the theoretical (solid and dashed lines in the left half of every column) and experimental (dotted lines in the right half of every column) emission energies from $X^{+}$ to $X^{2-}$. (d) Depiction of the initial and final states of each PL emission line. \label{LQDM1F}} \end{center}
\end{figure*}

To explain the relative intensities of multiple $X^{0}$ and $X^{+}$ PL lines, we consider the dynamics of charge relaxation within the LQDM. Normally, when an e-h pair is optically excited, the electrons relax first and the hole follows, with Coulomb interactions driving the hole toward the same QD. The electron and hole can relax into either QD, but it is energetically favorable for them to relax into the lower energy (right) QD. By measuring the energy difference between the two $X^{0}$ PL lines shown in Figure~\ref{positiveF}(a) and Figure~\ref{LQDM1F}(a), we learn that the $X^{+}$ state of LQDM 1 is 1.6 meV below the $X^{0}$, which is relatively small. This nearly-degenerate structure enables the optically or electrically injected electrons to relax to both left and right dots when no charges occupy the LQDM. For applied voltage below V1, $X^{0}$ PL emission from both the left and right QDs are observed because the electron can relax into either QD and the hole will follow the electron into the same QD. Electron-hole pairs relax more often into the low energy QD (right), therefore, we see $\begin{pmatrix}0&1\\0&1\\\end{pmatrix}$ has stronger PL intensities than $\begin{pmatrix}1&0\\1&0\\\end{pmatrix}$. Similarly, if the optically-excited electron tunnels out of the QDs and leaves the hole behind, it is energetically favorable for the hole to relax to the bigger QD (right), which has a lower energy confined state. Although the hole tunneling is slower than electron tunneling, the hole lifetime is limited by tunneling escape from the LQDM, not radiative recombination, and thus there is significant opportunity for the hole to relax. The presence of this hole in the right QD drives the electron to localize into the right QD. Consequently, we do not see emission from a positive trion state in which the electron is predominantly located in the left QD (includes $\begin{pmatrix}1&0\\1&1\\\end{pmatrix}$ and $\begin{pmatrix}1&0\\2&0\\\end{pmatrix}$). The optically excited hole will relax into the left QD because of hole-hole repulsion and serve as a spectator charge of the PL emission $\begin{pmatrix}0&1\\1&1\\\end{pmatrix}_{D}$ as a direct recombination. As is discussed previously, the interdot tunneling of the single electron  allows weak indirect recombination and emitted PL signal marked as $\begin{pmatrix}0&1\\1&1\\\end{pmatrix}_{I}$. We observe emission involving both positive trion and neutral excitons in the bias region to the left of V1 because the CCD integrates over multiple optical excitation and emission cycles that randomly contain both $X^{+}$ and $X^{0}$ events. The relative probability of these events is influenced by the laser power density as described above.

\section{Charging sequence in nearly degenerate LQDMs}\label{LQDM 1}
\subsection{$X^{-}$ charge configurations}
As the applied voltage moves the confined states of the QDs past the Fermi level, additional electrons tunnel into the LQDM.  This leads to a sequence of charging events and discrete shifts in the energies of ground state PL emission, as shown in Figure 3a. In order to understand the mechanism and consequences of this charging sequence in LQDMs, we systemically study the ground state spectra of LQDM 1, as a function of increasing applied voltage, by compare the experimental PL signatures with the theoretical modeling results. The charge configuration for almost all PL lines in Fig.~\ref{LQDM1F}(a) are assigned and indicated by the inset labels. We now justify and explain these assignments.

The first observation is that there are four values of the applied voltage (V1 through V4 at which discrete shifts in the PL spectra occur. These shifts occur as the increasing vertical bias moves confined energy states . The increasing of the vertical bias drives the conduction band across the Fermi level and electrically injects electrons one by one as is shown in Fig.~\ref{LQDM1F}(b).

The first group of PL lines, appearing for voltages above V1, are assigned to $\begin{pmatrix}1&1\\1&0\\\end{pmatrix}$, $\begin{pmatrix}1&1\\0&1\\\end{pmatrix}$, and$\begin{pmatrix}0&2\\0&1\\\end{pmatrix}$, all of which have two  electrons and one hole. In each of these configurations at least one electron is in the right QD. The second, optically generated, electron can relax into either the left or right QD. The hole, which typically relaxes more slowly, can also relax into either QD. We do not observe any lines assigned to $\begin{pmatrix}0&2\\1&0\\\end{pmatrix}$ because Coulomb attraction makes it unlikely that the hole will relax into a QD with no electrons. The PL emission of
$\begin{pmatrix}2&0\\1&0\\\end{pmatrix}$ can only happen when both the electrically-injected electron and optically-generated electron-hole pair occupy the left QD. Because the conduction level of the left QD is a little higher than the right one, the electrically-generated electron can be injected into the left QD only at voltages slightly higher than V1.

At the same time that the $X^{-}$ PL lines appear at V1, the $X^{+}$ PL line $\begin{pmatrix}0&1\\1&1\\\end{pmatrix}$ disappears. This occurs because electrical charging of the LQDM with a single electron makes it impossible for a single optically-generated hole to remain in the LQDM. For voltages larger than V1, the PL line $\begin{pmatrix}1&0\\1&0\\\end{pmatrix}$ gains significant optical intensity. This occurs because the probability that an electron relaxes to the higher energy (left) QD increases in the absence of a single hole in the right QD. Both neutral exciton states are observed at voltages above V1, despite the expectation that the LQDM should be charged with an excess electron. This is because the relaxation of the optically-generated electron into the LQDM can be blocked by both Coulomb and the Pauli interactions with the electron/s already occupying the LQDM. This relaxation blockade can force the electron to remain, temporarily, in a higher energy confined state from which tunneling out of the LQDM is more probable. It is therefore possible to observe PL emission of both charge states near the charging point. Similar processes lead to overlap of emission from other total charge states.\cite{Baier2001}.

Fig.~\ref{LQDM1F}(c) compares experimental (dotted lines) and theoretical (solid and dashed lines) PL energies for different excitonic complexes. The dashed lines in $X^{+}$ column represent the PL emissions that are energetically unfavorable, as is discussed in last section.
Good agreement is found for $X^+$ and $X^0$, as well as for $X^-$ complexes containing all charges within the same QD.
However, for $X^-$ with one electron in each QD in initial states, the theory predicts that $\begin{pmatrix}1&1\\1&0\\\end{pmatrix}$ and $\begin{pmatrix}1&1\\0&1\\\end{pmatrix}$
emission energies are blue and red shifted with respect to the corresponding $X^0$ levels. This is because of the participation of electrons
inter-dot tunnel coupling in the final state\cite{Royo2011}.
By contrast, in the experiment only a slight red shift of both lines is observed, similar to what is seen in VQDMs PL when spectator electrons are placed in neighbor QDs. This result lead us to suspect that the electron tunnel coupling in LQDM 1 may be suppressed for $X^{-}$ states. This suppression may be related to the different nature of repulsive Coulomb interactions between electron-electron and hole-hole. The basin structure in between the QD pair could also influence the tunneling strength.\cite{Peng2010} The simulations also predict the inverted energy of $\begin{pmatrix}0&2\\0&1\\\end{pmatrix}$ and $\begin{pmatrix}2&0\\1&0\\\end{pmatrix}$ emissions as a result of the inter-dot tunneling of the electron in the final state \cite{Munoz2011}. Our experimental techniques do not allow us to confirm this assignment, but it seems consistent with the fact that former shows up at smaller bias than the latter in Fig.~\ref{LQDM1F}(a).

\subsection{$X^{2-}$ charge configurations}
As the applied voltage increases beyond V1, second electron can tunnel into the LQDM. For LQDM 1, shown in Fig.~\ref{LQDM1F}(b), the confined conduction band energy levels of the two QDs are similar in energy and it is energetically favorable for the second electron to go into the left QD. Although the left QD energy state is at somewhat higher energy than the state of the right QD, this spatial configuration reduces the energy penalty of on-site electron-electron Coulomb repulsion if both electrons are in the right QD. As a result of the electrical injection of a second electron, new PL lines appear for voltages above V2. We assign these PL lines to the$\begin{pmatrix}1&2\\0&1\\\end{pmatrix}$ and $\begin{pmatrix}2&1\\1&0\\\end{pmatrix}$ charge configurations. We observe fine structure in these lines that we attribute to singlet and triplet electron spin configurations in the final state after optical recombination.

As is shown in the $X^{2-}$ column of Fig.~\ref{LQDM1F}(c), the experimental PL lines are a few meV lower than the simulated PL energies. Yet, the 0.4 meV energy splitting between $\begin{pmatrix}2&1\\1&0\\\end{pmatrix}$ and $\begin{pmatrix}1&2\\0&1\\\end{pmatrix}$ emission observed in the data agrees well with the computational prediction.

\subsection{$X^{3-}$ and higher charge configurations}
Continuing to increase the voltage makes it possible for a third electron to tunnel into the LQDM. As a result, $X^{3-}$ charge configurations become visible for voltages larger than V3. In the $X^{3-}$ charge configuration, no spin fine structure is expected to be possible. Three lines and a hint of the fourth line are visible, corresponding to the combination of two initial states and two final states of $X^{3-}$ emission. The small energy separation between each line suggests the perturbation from electron tunnel coupling and the Auger process in between two neighbor QDs.

As the applied voltage increase beyond V4, the ground state of the LQDM is filled by four electrically-charged electrons.  Optically generated electrons therefore occupy excited states of the LQDM. These excited states are delocalized over the entire LQDM and there is a relatively small energy spacing between each excited state. Consequently, we observe a quasi-continuous stark shift of the PL line as increasing numbers of electrons occupy the excited states\cite{Zhou2011, Peng2010}.

To summarize this section, we have provided detailed understanding of the charging process of LQDMs with nearly degenerate QDs by comparing experimental PL data with theoretical estimates and logical relaxation dynamics.  The observation of energy shifts computationally predicted to arise from electron tunneling provides strong experimental evidence for the existence of tunnel-coupling for $X^+$ (and possibly $X^-$) in this system.

\section{Charging sequence in non-degenerate LQDMs}
\label{LQDM2}
The detailed analysis presented in the previous section was possible because LQDM 1 happens to have QDs that are nearly degenerate in energy. Because of the self-assembly of LQDMs involves diffusion, the geometrical structure and composition profile can vary significantly between LQDMs. Consequently, many LQDMs with non-degenerate energy states are expected to be present in the sample. We present one example of a non-degenerate LQDM and show how the relaxation dynamics and charge interaction signatures developed in the previous section can be used to assign observed PL lines that appear dramatically different from the degenerate LQDM case.

\begin{figure}[htb]
\begin{center}
\includegraphics[width=8.0cm]{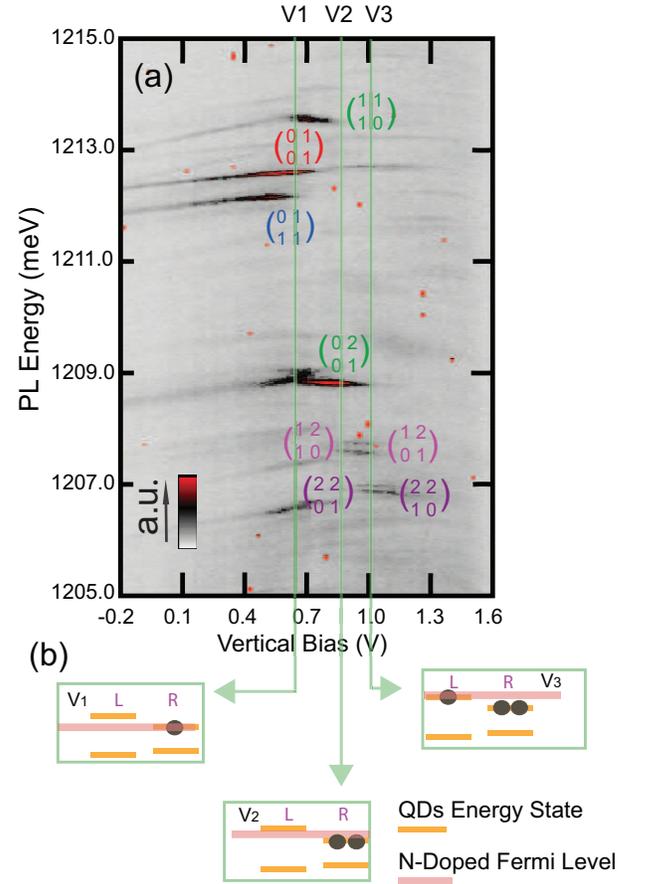}
\caption{(Color Online) (a)Full PL spectra of the ground states transition mapping as a function of vertical voltage measured for LQDM 2. (b) Electrically charging sequence as the increasing of the applied voltage. \label{LQDM2F}} \end{center}
\end{figure}

Fig.~\ref{LQDM2F}(a) shows the PL signature of LQDM 2 as function of vertical field. Two PL lines separated by 0.5 meV are observed for negative bias (1212.1 and 1212.6 meV at 0.5V). These PL lines are assigned to the $\begin{pmatrix}0&1\\0&1\\\end{pmatrix}$ and $\begin{pmatrix}0&1\\1&1\\\end{pmatrix}$ states following the discussion of positive trion states presented above. Similar to LQDM 1, we can clearly see discrete energy shifts due to Coulomb interactions when the vertical voltage allows one, two and three additional electrons to tunnel into the LQDMs at V1, V2 and V3. The locations of these charges, as shown in Fig.~\ref{LQDM2F}(b), are different from the degenerate LQDM case because there is substantially larger offset between the confined energy states of the left and right QD. We followed the same model as is described in last section to assign the different charge states of each PL emission as is marked on Fig.~\ref{LQDM2F}(a), the probability of relaxing into the high energy QD much smaller. We describe the significant differences here.

First, for voltage below V1, only one PL emission from a $X^{0}$ state is observed $\begin{pmatrix}0&1\\0&1\\\end{pmatrix}$. PL line $\begin{pmatrix}1&0\\1&0\\\end{pmatrix}$ , which originates in the emission of the neutral exciton in the higher energy QD is not observed because the optically excited electrons and holes relax to the low-energy QD faster than optical recombination can occur.

Second, only a single $X^{+}$ emission is observed in LQDM 2. As we discussed in the last two sections, the PL doublet for $\begin{pmatrix}0&1\\1&1\\\end{pmatrix}$ comes from different final states of the hole after electron-hole recombination. In a nearly-degenerate LQDM, a single electron is able to tunnel between two neighboring QDs when there is one hole in each QD. However, in a non-degenerate LQDM, electron tunneling is significantly suppressed and it is impossible for electrons in the right QD to recombine with the hole in the left QD. Therefore, only one emission line is observed in $X^{+}$ state.

Third, the PL line $\begin{pmatrix}1&1\\0&1\\\end{pmatrix}$, which is observed in degenerate LQDMs when the conduction level of the right (higher energy) QD crosses the Fermi level, is missing in this system. In the non-degenerate LQDM the PL emission from $\begin{pmatrix}1&1\\1&0\\\end{pmatrix}$ and $\begin{pmatrix}0&2\\0&1\\\end{pmatrix}$ turns on at this electric field. This change is a consequence of the offset in confined energy states in the non-degenerate LQDM. When one electrically-injected electron has occupied the conduction level of the low-energy QD, the optically generated electron will relax to the energetically favorable QD. In this case, the energy difference between the two QDs is relatively large, which drives the second electron to relax into the QD already occupied by one electron regardless of the Coulomb repulsion. The hole follows the electrons into the right QD and consequently beyond V1, $\begin{pmatrix}0&2\\0&1\\\end{pmatrix}$ is emitted and $\begin{pmatrix}1&1\\0&1\\\end{pmatrix}$ is missing. If the electron relaxes into the left QD, it creates a metastable state that rapidly relaxes to two electrons in the right QD. If, however, the hole follows the electron into the left QD, the hole is trapped by its large effective mass and slow tunneling. As a result, the Coulomb binding energy extends the lifetime of this metastable state and PL emission from the $\begin{pmatrix}1&1\\1&0\\\end{pmatrix}$ state can be observed.

 Beyond V2, the applied voltage charges LQDM 2 with two electrons. This charging results in a red shift by 1 meV between PL lines $\begin{pmatrix}0&2\\0&1\\\end{pmatrix}$ and the $X^{2-}$ states $\begin{pmatrix}1&2\\0&1\\\end{pmatrix}$ and $\begin{pmatrix}1&2\\1&0\\\end{pmatrix}$. We observe two discrete PL lines for the $X^{2-}$ state because the two electrically injected electrons relax into the right QD. The optically generated electron relaxes into the left QD rather than occupying an excited state of the right QD. The presence of electrons in both QDs makes it possible for the hole to relax into either QD.

 At V3, the electrical injection of the third electron causes a 0.8 meV red shift for the $X^{3-}$ PL emission. Again, a PL doublet is observed, corresponding to the relaxation of the hole into the left or right QD. The intensities of $X^{2-}$ and $X^{3-}$ PL lines are substantially reduced compared to $X^{-}$ PL lines. We observe significant variation between LQDMs in the PL intensities of highly charged states. We tentatively assign this variation to changes in electron tunneling escape rates from higher energy states, but further work to analyze this effect is necessary.

\section{Conclusion}
\label{Conclusion}
We analyze the PL emission of LQDMs as they are controllably charged with electrons in a vertical electric field geometry. Using laser intensity and electric field dependent measurements, along with an analysis of relaxation dynamics, we assign the observed PL emission to specific charge configurations. The observed energies are found to be in good quantitative agreement with calculations. The results reveal that changes in the relative energy of the two QDs comprising the LQDM can have substantial impact on the resulting PL spectral maps. Our investigation of an LQDM with nearly-degenerate energy levels, in conjunction with computational analysis, demonstrates the existence of inter-dot tunneling of electrons. For the non-degenerate LQDM, the evidence of inter-dot tunneling is not observed in zero lateral bias. Future work will focus on applying lateral electric field to simultaneously control the degeneracy of LQDMs.

\section{Acknowledgement}
This work was financially supported by NSF DMR-0844747 for Zhou/Doty, the Basic Science Program through the National Research Foundation (NRF) of Korea funded by the Ministry of Education, Science and Technology (Grant Nos. 2010-0008394 and 2011-0030821) for Jihoon Lee and MICINN project CTQ2011-27324 for JIC.


\begin{thebibliography}{27}
\expandafter\ifx\csname natexlab\endcsname\relax\def\natexlab#1{#1}\fi
\expandafter\ifx\csname bibnamefont\endcsname\relax
  \def\bibnamefont#1{#1}\fi
\expandafter\ifx\csname bibfnamefont\endcsname\relax
  \def\bibfnamefont#1{#1}\fi
\expandafter\ifx\csname citenamefont\endcsname\relax
  \def\citenamefont#1{#1}\fi
\expandafter\ifx\csname url\endcsname\relax
  \def\url#1{\texttt{#1}}\fi
\expandafter\ifx\csname urlprefix\endcsname\relax\def\urlprefix{URL }\fi
\providecommand{\bibinfo}[2]{#2}
\providecommand{\eprint}[2][]{\url{#2}}

\bibitem[{\citenamefont{Doty et~al.}(2010)\citenamefont{Doty, Climente,
  Greilich, Yakes, Bracker, and Gammon}}]{Doty2010}
\bibinfo{author}{\bibfnamefont{M.~F.} \bibnamefont{Doty}},
  \bibinfo{author}{\bibfnamefont{J.~I.} \bibnamefont{Climente}},
  \bibinfo{author}{\bibfnamefont{a.}~\bibnamefont{Greilich}},
  \bibinfo{author}{\bibfnamefont{M.}~\bibnamefont{Yakes}},
  \bibinfo{author}{\bibfnamefont{a.~S.} \bibnamefont{Bracker}},
  \bibnamefont{and} \bibinfo{author}{\bibfnamefont{D.}~\bibnamefont{Gammon}},
  \bibinfo{journal}{Journal of Physics: Conference Series}
  \textbf{\bibinfo{volume}{245}} (\bibinfo{year}{2010}).

\bibitem[{\citenamefont{Liu et~al.}(2010)\citenamefont{Liu, Bracker, Gammon,
  and Doty}}]{Liu2010}
\bibinfo{author}{\bibfnamefont{W.}~\bibnamefont{Liu}},
  \bibinfo{author}{\bibfnamefont{A.}~\bibnamefont{Bracker}},
  \bibinfo{author}{\bibfnamefont{D.}~\bibnamefont{Gammon}}, \bibnamefont{and}
  \bibinfo{author}{\bibfnamefont{M.~F.} \bibnamefont{Doty}},
  \bibinfo{journal}{in review}  (\bibinfo{year}{2010}).

\bibitem[{\citenamefont{Vamivakas et~al.}(2010)\citenamefont{Vamivakas, Lu,
  Matthiesen, Zhao, Falt, Badolato, and Atature}}]{Vamivakas2010}
\bibinfo{author}{\bibfnamefont{A.}~\bibnamefont{Vamivakas}},
  \bibinfo{author}{\bibfnamefont{C.}~\bibnamefont{Lu}},
  \bibinfo{author}{\bibfnamefont{C.}~\bibnamefont{Matthiesen}},
  \bibinfo{author}{\bibfnamefont{Y.}~\bibnamefont{Zhao}},
  \bibinfo{author}{\bibfnamefont{S.}~\bibnamefont{Falt}},
  \bibinfo{author}{\bibfnamefont{A.}~\bibnamefont{Badolato}}, \bibnamefont{and}
  \bibinfo{author}{\bibfnamefont{M.}~\bibnamefont{Atature}},
  \bibinfo{journal}{Nature} \textbf{\bibinfo{volume}{467}},
  \bibinfo{pages}{297} (\bibinfo{year}{2010}).

\bibitem[{\citenamefont{Wang et~al.}(2009)\citenamefont{Wang, Rastelli,
  Kiravittaya, Benyoucef, and Schmidt}}]{Wang2009}
\bibinfo{author}{\bibfnamefont{L.~J.} \bibnamefont{Wang}},
  \bibinfo{author}{\bibfnamefont{A.}~\bibnamefont{Rastelli}},
  \bibinfo{author}{\bibfnamefont{S.}~\bibnamefont{Kiravittaya}},
  \bibinfo{author}{\bibfnamefont{M.}~\bibnamefont{Benyoucef}},
  \bibnamefont{and} \bibinfo{author}{\bibfnamefont{O.~G.}
  \bibnamefont{Schmidt}}, \bibinfo{journal}{Advanced Materials}
  \textbf{\bibinfo{volume}{21}}, \bibinfo{pages}{2601} (\bibinfo{year}{2009}).

\bibitem[{\citenamefont{Kim et~al.}(2010)\citenamefont{Kim, Carter, Greilich,
  Bracker, and Gammon}}]{Kim2010b}
\bibinfo{author}{\bibfnamefont{D.}~\bibnamefont{Kim}},
  \bibinfo{author}{\bibfnamefont{S.}~\bibnamefont{Carter}},
  \bibinfo{author}{\bibfnamefont{A.}~\bibnamefont{Greilich}},
  \bibinfo{author}{\bibfnamefont{A.}~\bibnamefont{Bracker}}, \bibnamefont{and}
  \bibinfo{author}{\bibfnamefont{D.}~\bibnamefont{Gammon}},
  \bibinfo{journal}{Nature Physics}  (\bibinfo{year}{2010}).

\bibitem[{\citenamefont{Economou et~al.}(2012)\citenamefont{Economou, Climente,
  Badolato, Bracker, Gammon, and Doty}}]{Economou2012}
\bibinfo{author}{\bibfnamefont{S.~E.} \bibnamefont{Economou}},
  \bibinfo{author}{\bibfnamefont{J.~I.} \bibnamefont{Climente}},
  \bibinfo{author}{\bibfnamefont{A.}~\bibnamefont{Badolato}},
  \bibinfo{author}{\bibfnamefont{A.~S.} \bibnamefont{Bracker}},
  \bibinfo{author}{\bibfnamefont{D.}~\bibnamefont{Gammon}}, \bibnamefont{and}
  \bibinfo{author}{\bibfnamefont{M.~F.} \bibnamefont{Doty}},
  \bibinfo{journal}{Phys. Rev. B} \textbf{\bibinfo{volume}{86}},
  \bibinfo{pages}{085319} (\bibinfo{year}{2012}),
  \urlprefix\url{http://link.aps.org/doi/10.1103/PhysRevB.86.085319}.

\bibitem[{\citenamefont{Krenner et~al.}(2005)\citenamefont{Krenner, Sabathil,
  Clark, Kress, Schuh, Bichler, Abstreiter, and Finley}}]{Krenner2005}
\bibinfo{author}{\bibfnamefont{H.~J.} \bibnamefont{Krenner}},
  \bibinfo{author}{\bibfnamefont{M.}~\bibnamefont{Sabathil}},
  \bibinfo{author}{\bibfnamefont{E.~C.} \bibnamefont{Clark}},
  \bibinfo{author}{\bibfnamefont{A.}~\bibnamefont{Kress}},
  \bibinfo{author}{\bibfnamefont{D.}~\bibnamefont{Schuh}},
  \bibinfo{author}{\bibfnamefont{M.}~\bibnamefont{Bichler}},
  \bibinfo{author}{\bibfnamefont{G.}~\bibnamefont{Abstreiter}},
  \bibnamefont{and} \bibinfo{author}{\bibfnamefont{J.~J.}
  \bibnamefont{Finley}}, \bibinfo{journal}{Physical Review Letters}
  \textbf{\bibinfo{volume}{94}}, \bibinfo{pages}{57402} (\bibinfo{year}{2005}).

\bibitem[{\citenamefont{Stinaff et~al.}(2006)\citenamefont{Stinaff, Scheibner,
  Bracker, Ponomarev, Korenev, Ware, Doty, Reinecke, and Gammon}}]{Stinaff2006}
\bibinfo{author}{\bibfnamefont{E.~A.} \bibnamefont{Stinaff}},
  \bibinfo{author}{\bibfnamefont{M.}~\bibnamefont{Scheibner}},
  \bibinfo{author}{\bibfnamefont{A.~S.} \bibnamefont{Bracker}},
  \bibinfo{author}{\bibfnamefont{I.~V.} \bibnamefont{Ponomarev}},
  \bibinfo{author}{\bibfnamefont{V.~L.} \bibnamefont{Korenev}},
  \bibinfo{author}{\bibfnamefont{M.~E.} \bibnamefont{Ware}},
  \bibinfo{author}{\bibfnamefont{M.~F.} \bibnamefont{Doty}},
  \bibinfo{author}{\bibfnamefont{T.~L.} \bibnamefont{Reinecke}},
  \bibnamefont{and} \bibinfo{author}{\bibfnamefont{D.}~\bibnamefont{Gammon}},
  \bibinfo{journal}{Science} \textbf{\bibinfo{volume}{311}},
  \bibinfo{pages}{636} (\bibinfo{year}{2006}).

\bibitem[{\citenamefont{Doty et~al.}(2006)\citenamefont{Doty, Scheibner,
  Ponomarev, Stinaff, Bracker, Korenev, Reinecke, and Gammon}}]{Doty2006}
\bibinfo{author}{\bibfnamefont{M.~F.} \bibnamefont{Doty}},
  \bibinfo{author}{\bibfnamefont{M.}~\bibnamefont{Scheibner}},
  \bibinfo{author}{\bibfnamefont{I.~V.} \bibnamefont{Ponomarev}},
  \bibinfo{author}{\bibfnamefont{E.~A.} \bibnamefont{Stinaff}},
  \bibinfo{author}{\bibfnamefont{A.~S.} \bibnamefont{Bracker}},
  \bibinfo{author}{\bibfnamefont{V.~L.} \bibnamefont{Korenev}},
  \bibinfo{author}{\bibfnamefont{T.~L.} \bibnamefont{Reinecke}},
  \bibnamefont{and} \bibinfo{author}{\bibfnamefont{D.}~\bibnamefont{Gammon}},
  \bibinfo{journal}{Physical Review Letters} \textbf{\bibinfo{volume}{97}},
  \bibinfo{pages}{197202} (\bibinfo{year}{2006}).

\bibitem[{\citenamefont{Doty et~al.}(2009)\citenamefont{Doty, Climente,
  Korkusinski, Scheibner, Bracker, Hawrylak, and Gammon}}]{Doty2009}
\bibinfo{author}{\bibfnamefont{M.~F.} \bibnamefont{Doty}},
  \bibinfo{author}{\bibfnamefont{J.~I.} \bibnamefont{Climente}},
  \bibinfo{author}{\bibfnamefont{M.}~\bibnamefont{Korkusinski}},
  \bibinfo{author}{\bibfnamefont{M.}~\bibnamefont{Scheibner}},
  \bibinfo{author}{\bibfnamefont{A.~S.} \bibnamefont{Bracker}},
  \bibinfo{author}{\bibfnamefont{P.}~\bibnamefont{Hawrylak}}, \bibnamefont{and}
  \bibinfo{author}{\bibfnamefont{D.}~\bibnamefont{Gammon}},
  \bibinfo{journal}{Physical Review Letters} \textbf{\bibinfo{volume}{102}},
  \bibinfo{pages}{47401} (\bibinfo{year}{2009}).

\bibitem[{\citenamefont{Ortner et~al.}(2005)\citenamefont{Ortner, Bayer,
  Geller, Reinecke, Kress, Reithmaier, and Forchel}}]{Ortner2005}
\bibinfo{author}{\bibfnamefont{G.}~\bibnamefont{Ortner}},
  \bibinfo{author}{\bibfnamefont{M.}~\bibnamefont{Bayer}},
  \bibinfo{author}{\bibfnamefont{Y.~L.} \bibnamefont{Geller}},
  \bibinfo{author}{\bibfnamefont{T.~L.} \bibnamefont{Reinecke}},
  \bibinfo{author}{\bibfnamefont{A.}~\bibnamefont{Kress}},
  \bibinfo{author}{\bibfnamefont{J.~P.} \bibnamefont{Reithmaier}},
  \bibnamefont{and} \bibinfo{author}{\bibfnamefont{A.}~\bibnamefont{Forchel}},
  \bibinfo{journal}{Physical Review Letters} \textbf{\bibinfo{volume}{94}},
  \bibinfo{pages}{157401} (\bibinfo{year}{2005}).

\bibitem[{\citenamefont{Ledentsov et~al.}(1996)\citenamefont{Ledentsov,
  Shchukin, Grundmann, Kirstaedter, Bohrer, Schmidt, Bimberg, Ustinov, Egorov,
  Zhukov et~al.}}]{Ledentsov1996}
\bibinfo{author}{\bibfnamefont{N.~N.} \bibnamefont{Ledentsov}},
  \bibinfo{author}{\bibfnamefont{V.~A.} \bibnamefont{Shchukin}},
  \bibinfo{author}{\bibfnamefont{M.}~\bibnamefont{Grundmann}},
  \bibinfo{author}{\bibfnamefont{N.}~\bibnamefont{Kirstaedter}},
  \bibinfo{author}{\bibfnamefont{J.}~\bibnamefont{Bohrer}},
  \bibinfo{author}{\bibfnamefont{O.~G.} \bibnamefont{Schmidt}},
  \bibinfo{author}{\bibfnamefont{D.}~\bibnamefont{Bimberg}},
  \bibinfo{author}{\bibfnamefont{V.~M.} \bibnamefont{Ustinov}},
  \bibinfo{author}{\bibfnamefont{A.~Y.} \bibnamefont{Egorov}},
  \bibinfo{author}{\bibfnamefont{A.~E.} \bibnamefont{Zhukov}},
  \bibnamefont{et~al.}, \bibinfo{journal}{Physical Review B}
  \textbf{\bibinfo{volume}{54}}, \bibinfo{pages}{8743} (\bibinfo{year}{1996}).

\bibitem[{\citenamefont{Lee et~al.}(2006)\citenamefont{Lee, Wang, Strom, Mazur,
  and Salamo}}]{Lee2006}
\bibinfo{author}{\bibfnamefont{J.~H.} \bibnamefont{Lee}},
  \bibinfo{author}{\bibfnamefont{Z.~M.} \bibnamefont{Wang}},
  \bibinfo{author}{\bibfnamefont{N.~W.} \bibnamefont{Strom}},
  \bibinfo{author}{\bibfnamefont{Y.~I.} \bibnamefont{Mazur}}, \bibnamefont{and}
  \bibinfo{author}{\bibfnamefont{G.~J.} \bibnamefont{Salamo}},
  \bibinfo{journal}{Applied Physics Letters} \textbf{\bibinfo{volume}{89}},
  \bibinfo{pages}{202101} (\bibinfo{year}{2006}).

\bibitem[{\citenamefont{Liang et~al.}(2006)\citenamefont{Liang, Wang, Lee,
  Sablon, Mazur, and Salamo}}]{Liang2006}
\bibinfo{author}{\bibfnamefont{B.~L.} \bibnamefont{Liang}},
  \bibinfo{author}{\bibfnamefont{Z.~M.} \bibnamefont{Wang}},
  \bibinfo{author}{\bibfnamefont{J.~H.} \bibnamefont{Lee}},
  \bibinfo{author}{\bibfnamefont{K.}~\bibnamefont{Sablon}},
  \bibinfo{author}{\bibfnamefont{Y.~I.} \bibnamefont{Mazur}}, \bibnamefont{and}
  \bibinfo{author}{\bibfnamefont{G.~J.} \bibnamefont{Salamo}},
  \bibinfo{journal}{Applied Physics Letters} \textbf{\bibinfo{volume}{89}},
  \bibinfo{pages}{043113} (\bibinfo{year}{2006}).

\bibitem[{\citenamefont{Liang et~al.}({2008})\citenamefont{Liang, Wang, Wang,
  Lee, Mazur, Shih, and Salamo}}]{Liang2008}
\bibinfo{author}{\bibfnamefont{B.~L.} \bibnamefont{Liang}},
  \bibinfo{author}{\bibfnamefont{Z.~M.} \bibnamefont{Wang}},
  \bibinfo{author}{\bibfnamefont{X.~Y.} \bibnamefont{Wang}},
  \bibinfo{author}{\bibfnamefont{J.~H.} \bibnamefont{Lee}},
  \bibinfo{author}{\bibfnamefont{Y.~I.} \bibnamefont{Mazur}},
  \bibinfo{author}{\bibfnamefont{C.~K.} \bibnamefont{Shih}}, \bibnamefont{and}
  \bibinfo{author}{\bibfnamefont{G.~J.} \bibnamefont{Salamo}},
  \bibinfo{journal}{{ACS NANO}}  (\bibinfo{year}{{2008}}).

\bibitem[{\citenamefont{Peng et~al.}(2010)\citenamefont{Peng, Hermannstadter,
  Witzany, Heldmaier, Wang, Kiravittaya, Rastelli, Schmidt, Michler, and
  Bester}}]{Peng2010}
\bibinfo{author}{\bibfnamefont{J.}~\bibnamefont{Peng}},
  \bibinfo{author}{\bibfnamefont{C.}~\bibnamefont{Hermannstadter}},
  \bibinfo{author}{\bibfnamefont{M.}~\bibnamefont{Witzany}},
  \bibinfo{author}{\bibfnamefont{M.}~\bibnamefont{Heldmaier}},
  \bibinfo{author}{\bibfnamefont{L.}~\bibnamefont{Wang}},
  \bibinfo{author}{\bibfnamefont{S.}~\bibnamefont{Kiravittaya}},
  \bibinfo{author}{\bibfnamefont{A.}~\bibnamefont{Rastelli}},
  \bibinfo{author}{\bibfnamefont{O.~G.} \bibnamefont{Schmidt}},
  \bibinfo{author}{\bibfnamefont{P.}~\bibnamefont{Michler}}, \bibnamefont{and}
  \bibinfo{author}{\bibfnamefont{G.}~\bibnamefont{Bester}},
  \bibinfo{journal}{Physical Review B} \textbf{\bibinfo{volume}{81}},
  \bibinfo{pages}{205315} (\bibinfo{year}{2010}).

\bibitem[{\citenamefont{Royo et~al.}({2011})\citenamefont{Royo, Climente, and
  Planelles}}]{Royo2011}
\bibinfo{author}{\bibfnamefont{M.}~\bibnamefont{Royo}},
  \bibinfo{author}{\bibfnamefont{J.~I.} \bibnamefont{Climente}},
  \bibnamefont{and}
  \bibinfo{author}{\bibfnamefont{J.}~\bibnamefont{Planelles}},
  \bibinfo{journal}{{PHYSICAL REVIEW B}} \textbf{\bibinfo{volume}{{84}}}
  (\bibinfo{year}{{2011}}), ISSN \bibinfo{issn}{{1098-0121}}.

\bibitem[{\citenamefont{Zhou et~al.}({2011})\citenamefont{Zhou, Sanwlani, Liu,
  Lee, Wang, Salamo, and Doty}}]{Zhou2011}
\bibinfo{author}{\bibfnamefont{X.}~\bibnamefont{Zhou}},
  \bibinfo{author}{\bibfnamefont{S.}~\bibnamefont{Sanwlani}},
  \bibinfo{author}{\bibfnamefont{W.}~\bibnamefont{Liu}},
  \bibinfo{author}{\bibfnamefont{J.~H.} \bibnamefont{Lee}},
  \bibinfo{author}{\bibfnamefont{Z.~M.} \bibnamefont{Wang}},
  \bibinfo{author}{\bibfnamefont{G.}~\bibnamefont{Salamo}}, \bibnamefont{and}
  \bibinfo{author}{\bibfnamefont{M.~F.} \bibnamefont{Doty}},
  \bibinfo{journal}{{PHYSICAL REVIEW B}} \textbf{\bibinfo{volume}{{84}}}
  (\bibinfo{year}{{2011}}), ISSN \bibinfo{issn}{{1098-0121}}.

\bibitem[{\citenamefont{Munoz-Matutano
  et~al.}(2011)\citenamefont{Munoz-Matutano, Royo, Climente, Canet-Ferrer,
  Fuster, Alonso-Gonzalez, Fernandez-Martinez, Martinez-Pastor, Gonzalez,
  Gonzalez et~al.}}]{Munoz2011}
\bibinfo{author}{\bibfnamefont{G.}~\bibnamefont{Munoz-Matutano}},
  \bibinfo{author}{\bibfnamefont{M.}~\bibnamefont{Royo}},
  \bibinfo{author}{\bibfnamefont{J.~I.} \bibnamefont{Climente}},
  \bibinfo{author}{\bibfnamefont{J.}~\bibnamefont{Canet-Ferrer}},
  \bibinfo{author}{\bibfnamefont{D.}~\bibnamefont{Fuster}},
  \bibinfo{author}{\bibfnamefont{P.}~\bibnamefont{Alonso-Gonzalez}},
  \bibinfo{author}{\bibfnamefont{I.}~\bibnamefont{Fernandez-Martinez}},
  \bibinfo{author}{\bibfnamefont{J.}~\bibnamefont{Martinez-Pastor}},
  \bibinfo{author}{\bibfnamefont{Y.}~\bibnamefont{Gonzalez}},
  \bibinfo{author}{\bibfnamefont{L.}~\bibnamefont{Gonzalez}},
  \bibnamefont{et~al.}, \bibinfo{journal}{Phys. Rev. B}
  \textbf{\bibinfo{volume}{84}}, \bibinfo{pages}{041308}
  (\bibinfo{year}{2011}).

\bibitem[{\citenamefont{Heldmaier et~al.}({2012})\citenamefont{Heldmaier,
  Seible, Hermannstaedter, Witzany, Rossbach, Jetter, Michler, Wang, Rastelli,
  and Schmidt}}]{Heldmaier2012}
\bibinfo{author}{\bibfnamefont{M.}~\bibnamefont{Heldmaier}},
  \bibinfo{author}{\bibfnamefont{M.}~\bibnamefont{Seible}},
  \bibinfo{author}{\bibfnamefont{C.}~\bibnamefont{Hermannstaedter}},
  \bibinfo{author}{\bibfnamefont{M.}~\bibnamefont{Witzany}},
  \bibinfo{author}{\bibfnamefont{R.}~\bibnamefont{Rossbach}},
  \bibinfo{author}{\bibfnamefont{M.}~\bibnamefont{Jetter}},
  \bibinfo{author}{\bibfnamefont{P.}~\bibnamefont{Michler}},
  \bibinfo{author}{\bibfnamefont{L.}~\bibnamefont{Wang}},
  \bibinfo{author}{\bibfnamefont{A.}~\bibnamefont{Rastelli}}, \bibnamefont{and}
  \bibinfo{author}{\bibfnamefont{O.~G.} \bibnamefont{Schmidt}},
  \bibinfo{journal}{{PHYSICAL REVIEW B}} \textbf{\bibinfo{volume}{{85}}}
  (\bibinfo{year}{{2012}}), ISSN \bibinfo{issn}{{1098-0121}}.

\bibitem[{\citenamefont{Beirne et~al.}({2006})\citenamefont{Beirne,
  Hermannstadter, Wang, Rastelli, Schmidt, and Michler}}]{Beirne2006}
\bibinfo{author}{\bibfnamefont{G.}~\bibnamefont{Beirne}},
  \bibinfo{author}{\bibfnamefont{C.}~\bibnamefont{Hermannstadter}},
  \bibinfo{author}{\bibfnamefont{L.}~\bibnamefont{Wang}},
  \bibinfo{author}{\bibfnamefont{A.}~\bibnamefont{Rastelli}},
  \bibinfo{author}{\bibfnamefont{O.}~\bibnamefont{Schmidt}}, \bibnamefont{and}
  \bibinfo{author}{\bibfnamefont{P.}~\bibnamefont{Michler}},
  \bibinfo{journal}{{PHYSICAL REVIEW LETTERS}} \textbf{\bibinfo{volume}{{96}}}
  (\bibinfo{year}{{2006}}), ISSN \bibinfo{issn}{{0031-9007}}.

\bibitem[{\citenamefont{Lee et~al.}(2010)\citenamefont{Lee, Wang, Dorogan,
  Mazur, and Salamo}}]{Lee2010}
\bibinfo{author}{\bibfnamefont{J.}~\bibnamefont{Lee}},
  \bibinfo{author}{\bibfnamefont{Z.}~\bibnamefont{Wang}},
  \bibinfo{author}{\bibfnamefont{V.}~\bibnamefont{Dorogan}},
  \bibinfo{author}{\bibfnamefont{Y.}~\bibnamefont{Mazur}}, \bibnamefont{and}
  \bibinfo{author}{\bibfnamefont{G.}~\bibnamefont{Salamo}},
  \bibinfo{journal}{IEEE Transactions on Nanotechnology}
  \textbf{\bibinfo{volume}{9}}, \bibinfo{pages}{149} (\bibinfo{year}{2010}).

\bibitem{Dalgarno2008}
P. A. Dalgarno, J. M. Smith, J. McFarlane, B. D. Gerardot, K. Karrai, A. Badolato, P. M. Petroff, and R. J. Warburton,
Phys. Rev. B {\bf 77}, 245311 (2008).

\bibitem[{\citenamefont{Finley et~al.}({2004})\citenamefont{Finley, Sabathil,
  Vogl, Abstreiter, Oulton, Tartakovskii, Mowbray, Skolnick, Liew, Cullis
  et~al.}}]{Finley2004}
\bibinfo{author}{\bibfnamefont{J.}~\bibnamefont{Finley}},
  \bibinfo{author}{\bibfnamefont{M.}~\bibnamefont{Sabathil}},
  \bibinfo{author}{\bibfnamefont{P.}~\bibnamefont{Vogl}},
  \bibinfo{author}{\bibfnamefont{G.}~\bibnamefont{Abstreiter}},
  \bibinfo{author}{\bibfnamefont{R.}~\bibnamefont{Oulton}},
  \bibinfo{author}{\bibfnamefont{A.}~\bibnamefont{Tartakovskii}},
  \bibinfo{author}{\bibfnamefont{D.}~\bibnamefont{Mowbray}},
  \bibinfo{author}{\bibfnamefont{M.}~\bibnamefont{Skolnick}},
  \bibinfo{author}{\bibfnamefont{S.}~\bibnamefont{Liew}},
  \bibinfo{author}{\bibfnamefont{A.}~\bibnamefont{Cullis}},
  \bibnamefont{et~al.}, \bibinfo{journal}{{PHYSICAL REVIEW B}}
  \textbf{\bibinfo{volume}{{70}}} (\bibinfo{year}{{2004}}), ISSN
  \bibinfo{issn}{{1098-0121}}.

\bibitem[{\citenamefont{Doty et~al.}(2008)\citenamefont{Doty, Scheibner,
  Bracker, Ponomarev, Reinecke, and Gammon}}]{Doty2008}
\bibinfo{author}{\bibfnamefont{M.~F.} \bibnamefont{Doty}},
  \bibinfo{author}{\bibfnamefont{M.}~\bibnamefont{Scheibner}},
  \bibinfo{author}{\bibfnamefont{A.~S.} \bibnamefont{Bracker}},
  \bibinfo{author}{\bibfnamefont{I.~V.} \bibnamefont{Ponomarev}},
  \bibinfo{author}{\bibfnamefont{T.~L.} \bibnamefont{Reinecke}},
  \bibnamefont{and} \bibinfo{author}{\bibfnamefont{D.}~\bibnamefont{Gammon}},
  \bibinfo{journal}{Physical Review B} \textbf{\bibinfo{volume}{78}},
  \bibinfo{pages}{115316} (\bibinfo{year}{2008}).

\bibitem[{\citenamefont{Baier et~al.}({2001})\citenamefont{Baier, Findeis,
  Zrenner, Bichler, and Abstreiter}}]{Baier2001}
\bibinfo{author}{\bibfnamefont{M.}~\bibnamefont{Baier}},
  \bibinfo{author}{\bibfnamefont{F.}~\bibnamefont{Findeis}},
  \bibinfo{author}{\bibfnamefont{A.}~\bibnamefont{Zrenner}},
  \bibinfo{author}{\bibfnamefont{M.}~\bibnamefont{Bichler}}, \bibnamefont{and}
  \bibinfo{author}{\bibfnamefont{G.}~\bibnamefont{Abstreiter}},
  \bibinfo{journal}{{PHYSICAL REVIEW B}} \textbf{\bibinfo{volume}{{64}}}
  (\bibinfo{year}{{2001}}), ISSN \bibinfo{issn}{{1098-0121}}.

\bibitem[{\citenamefont{Ediger et~al.}(2007)\citenamefont{Ediger, Bester,
  Badolato, Petroff, Karrai, Zunger, and Warburton}}]{Ediger2007a}
\bibinfo{author}{\bibfnamefont{M.}~\bibnamefont{Ediger}},
  \bibinfo{author}{\bibfnamefont{G.}~\bibnamefont{Bester}},
  \bibinfo{author}{\bibfnamefont{A.}~\bibnamefont{Badolato}},
  \bibinfo{author}{\bibfnamefont{P.~M.} \bibnamefont{Petroff}},
  \bibinfo{author}{\bibfnamefont{K.}~\bibnamefont{Karrai}},
  \bibinfo{author}{\bibfnamefont{A.}~\bibnamefont{Zunger}}, \bibnamefont{and}
  \bibinfo{author}{\bibfnamefont{R.~J.} \bibnamefont{Warburton}},
  \bibinfo{journal}{Nature Physics} \textbf{\bibinfo{volume}{3}},
  \bibinfo{pages}{774} (\bibinfo{year}{2007}).

\bibitem[{\citenamefont{Scheibner et~al.}(2007)\citenamefont{Scheibner,
  Ponomarev, Stinaff, Doty, Bracker, Hellberg, Reinecke, and
  Gammon}}]{Scheibner2007a}
\bibinfo{author}{\bibfnamefont{M.}~\bibnamefont{Scheibner}},
  \bibinfo{author}{\bibfnamefont{I.~V.} \bibnamefont{Ponomarev}},
  \bibinfo{author}{\bibfnamefont{E.~A.} \bibnamefont{Stinaff}},
  \bibinfo{author}{\bibfnamefont{M.~F.} \bibnamefont{Doty}},
  \bibinfo{author}{\bibfnamefont{A.~S.} \bibnamefont{Bracker}},
  \bibinfo{author}{\bibfnamefont{C.~S.} \bibnamefont{Hellberg}},
  \bibinfo{author}{\bibfnamefont{T.~L.} \bibnamefont{Reinecke}},
  \bibnamefont{and} \bibinfo{author}{\bibfnamefont{D.}~\bibnamefont{Gammon}},
  \bibinfo{journal}{Physical Review Letters} \textbf{\bibinfo{volume}{99}},
  \bibinfo{pages}{197402} (\bibinfo{year}{2007}).

\end{thebibliography}

\end{document}